\documentclass[16pt]{article}
\usepackage{amsmath}
\usepackage{amsfonts}
\usepackage{amssymb}
\usepackage{graphicx}
\usepackage{color} 
\usepackage{float}
\begin{document}
\title{Gravity Induced Resonant Emission}
\author{ Antony Soosaleon$^{1,2}$  Sunitha Amborse$^3$ \\
\small 1. The Astronomical Observatory, Department of Physics
\small University of Kerala,  India - 695033 \\
\small 2. SPAP, M.G.University, Kottayam,
\small Kerala, India-686560, antonysoosaleon@yahoo.com \\
\small 3. Center for Fundamental Research and Computational Sciences, \\
\small Thiruvanathapuram, Kerala, India - 695043. sunithaambrose@gmail.com}
\maketitle
\begin{abstract}
The gravitational drift of ions relative to the electrons induces two type of waves in magnetized plasma; ion acoustic (\emph{ia}) waves and lower hybrid (\emph{lh}) waves. The \emph{ia} waves induced by the gravity are damped by electromagnetic (\emph{em}) waves, which leads to the formation of \emph{lh} waves. For higher wave vector, these \emph{lh} waves results in to the resonant absorption \& re-emission of \emph{em} waves, called as gravity induced resonant emission (\emph{gire}). A general formula has been derived for \emph{gire} frequency, which is given by $\omega_r^2 = (\frac{c^2e^2B^2}{mMg})(\frac{n_0^{'}}{n_0})$; c, velocity of light; g, gravity; e, charge; B, magnetic field; M \& m and $ n_0^{'}$ \&  ${n_0}$ are masses and number densities of ion and electron respectively. 
\end{abstract}
\section{Introduction} 
Gravity is inherent, any fundamental theory including the effect of gravity would be more realistic. The effect of gravity in magnetized plasma had been discussed by Chen\cite{ref6} and Treumann \& Baumjohann\cite{ref19} as Rayleigh Taylor Instability, which did not yield any result. The problem with their study was that they considered the plasma as cold; for any instability, a driving force alone is not sufficient but it needs energy to support. Even otherwise too, plasma is an ionized state of matter exists only at high temperature and does not sustain with out magnetic field and therefore these two quantities are inevitable, by these reasons the plasma considered for this study is warm and magnetized. The effect of these two parameters is so volatile, hence they are set as finite for simplicity. The ionized particles in magnetic field undergo cyclotron motion, which is controlled by several factors like mass ($m_i$) \& mass number $\mu$, charge state (Ze) and the effective field strength (B) (cyclotron frequency, $\Omega_i \pm\frac{ZeB}{\mu m_i}$). The effective magnetic field on each ions is unique; even if we assume all the ions in a plasma experience the same magnetic field, the masses are different, the charge state often changes and the direction of the cyclotron motion is different for electrons and ions. With all these variations including the temperature, a magnetized warm plasma may possess a variety of localized fields and such fields can cause localized currents; sometimes certain field resonates with the other localized fields leads to several observable phenomenon. In an ultra vacuum condition (Solar Corona; \cite{ref2}), the currents by localized fields are not possible, rather the Coulomb's interaction will be dominant, by the fact that the fields will be effective to very long distances.  Here is a situation arises, that the gravitational drift of heavier particles relative to the electrons induces an electric field; this gravitationally induced electric field (\emph{gief}) couples with the magnetic field, causes two physical phenomenon. 1. It generates ion acoustic (\emph{ia}) waves and these waves are damped by the perpendicular propagation electromagnetic (\emph{em}) waves leads to the formation of lower hybrid waves 2. The gravity induced lower hybrid (\emph{lh}) waves leads to the resonant absorption and re-emission of \emph{em} waves and in both the cases, the Coulomb's force is playing a lead role. The \emph{gief} may not be effective to high $\beta$ plasmas ($\beta = \frac{\mu_{\circ}nKT}{B^2}$, is the ratio of particle pressure to the magnetic field pressure; e.g., plasmas in Fusion devises, stellar interior and photosphere) due to the high collisions.\\ This study is carried out by deriving a dispersion relation (relation between the angular frequency $\omega$ and the wave vector $k$) for the \emph{em} wave propagation in a warm magnetized gravitational plasma. We are considering only the perpendicular propagation ($\bot$ to B), by the reason that the Lorentz force is a normal force and hence the coupling is possible only in the perpendicular direction.
\section{Theoretical Description}
\subsection{Gravitational Plasma} The purpose of this study is to find the effect of gravitational force over the magnetized plasma, therefore  we need only ionized particles, for that we consider the plasma as fully ionized (coronal plasma). As briefed in the introduction, the ions are heavier, drift more relative to the electrons. The drift velocity of ions due to the gravity $\mathbf{v_0}$ is deduced from the general definition,  $\mathbf{v_f} = \frac{\mathbf{F}\times\mathbf{B}}{q B^2}$. For gravitational drift; $\mathbf{F}$ is replaced by $ m_i \mathbf{g}$, which gives $\mathbf{v_0} = (\frac{m_i}{q_i})\frac{\mathbf{g} \times \mathbf{B}}{ B^2} = -\frac{g}{\Omega_i}\hat{y}$
where $m_i$, mass; $q_i$, charge and g, acceleration due to gravity respectively.
\subsection{The Basic Equations Used} The dispersion relation is derived by using magneto hydro dynamical theory and the following equations are used. The momentum equation is given by
\begin{equation}
\begin{array}{l}
m_i n_i \frac{d{\mathbf{v_i}}}{d{t}}= q_i n_i [\mathbf{E}+(\mathbf{v_i} \times \mathbf{B})] + m_i n_i \mathbf{g}-\nabla (\gamma_i K T_i)   
\end{array}
\end{equation}
Here $n_i$, number density; $v_i$, velocity; $T_i$, temperature; $\gamma_i$, the ratio of specific heats;  E, electric field; K is the Boltzmann constant respectively. Since \emph{gief} is effective only to the low $ \beta $ plasmas, we have neglected the collision \& viscosity terms. The second equation is the continuity equation which is given by
\begin{equation} 
\frac{\partial{n_i}}{\partial{t}}+\nabla . (n_i \mathbf{v_i})=0
\end{equation}
\subsection{Procedure of the theory} 
\begin{enumerate}
	\item {The dispersion relation is derived by solving the momentum equation by linearization method.\cite{ref6}}
	\item {First we deduce a resultant perturbed momentum equation}
	\item {Then we introduce the \emph{em} waves as perturbation in the momentum equation and derive perturbed velocities}
	\item {Substituting these velocities in the perturbed continuity equation, the dispersion relation is derived}
\end{enumerate}
\begin{figure}
\centering
\includegraphics[height=4cm]{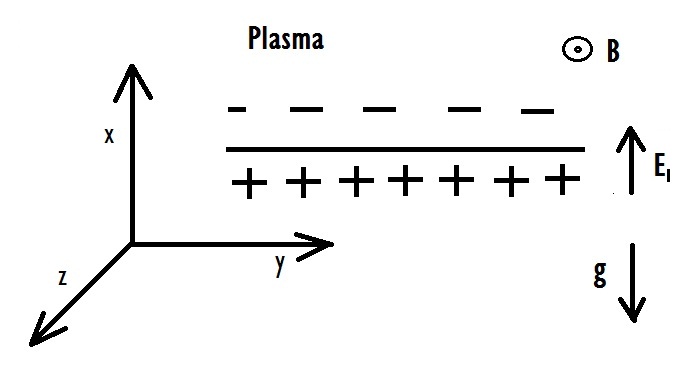}
\includegraphics[height=4cm]{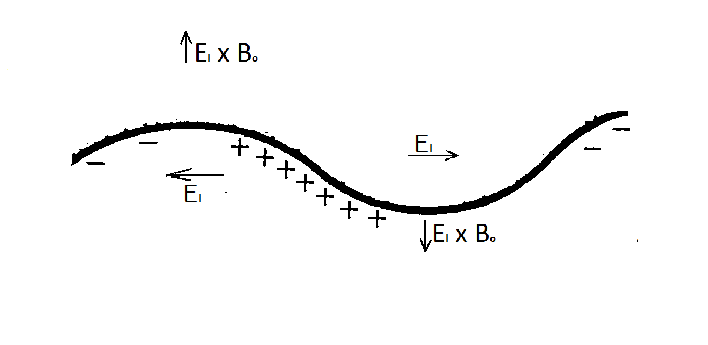}
\caption {Gravity induced electric field in unperturbed and perturbed magnetic field}
\end{figure}
\subsection{Derivation of Dispersion Relation}
Consider the gravitational force is acting in the x-axis, the wave is propagating in the y-axis and the magnetic field is applied in the z-axis as shown in the fig.1. The effect of gravity over the ions relative to the electrons causes two things; 1. a density gradient $\nabla n$ in the direction of x-axis and 2. it induces electric field $\mathbf{E_1}$ due to $\mathbf{v_0}$ in the -x axis. This induced field $\mathbf{E_1}$causes a drift $\mathbf{E_1} \times \mathbf{B}$ in the absence of perturbation and this drift couples with the diamagnetic drifts leads to formation of \emph{ia} oscillations. But this situation changes when there is a perturbation in the form of \emph{em} waves, the direction of $\mathbf{E_1}$ changes at every point and hence the direction of the drift $ \mathbf{E_1} \times \mathbf{B}$ also changes (see fig.2), which leads to the damping of \emph{ia} waves and the formation of \emph{lh} oscillation.
For simplicity, we are considering only a two component plasma (electrons \& ions), since the effect of gravity is over the density and the velocity (by the induced electric field), the theory is formulated by introducing the perturbation in the density $n_1$ and the velocity $\mathbf{v}_1$ of the ionized particles. First, we start with the ion component of the plasma; the perturbed momentum equation for ions is obtained, by replacing $n_i$ with $ n_0+n_1$ and $ {v}_i $  with $\mathbf{v}_0+\mathbf{v}_1$ in the Eq.(1), we get,
\begin{equation}
\begin{array}{l}
m_i(n_0+n_1) [\frac{\partial}{\partial{t}}(\mathbf{v}_0+\mathbf{v}_1) + (\mathbf{v}_0+\mathbf{v}_1).\nabla(\mathbf{v}_0+\mathbf{v}_1)]  = e ( n_0+ n_1)[\mathbf{E}_1+(\mathbf{v}_0+\mathbf{v}_1)\times \mathbf{B}_0]\\ + m_i(n_0+n_1)g - \gamma_i K T_i  \nabla n_1-\gamma_i K T_i  \nabla n_0
\end{array}
\end{equation}

The steady state momentum equation for ions is written as,
\begin{equation}
\begin{array}{l}
m_i n_0(\mathbf{v}_0 . \nabla)v_0=e n_0 (\mathbf{v}_0 \times \mathbf{B}_0)  + m_i n_0 \mathbf {g} - \gamma_i K T_i  \nabla n_0
\end{array} 
\end{equation}
Multiplying Eq.(4) with $(1+\frac{n_1}{n_0})$ and subtracting it from Eq.(3) and neglecting the second order terms, we get the resultant perturbed equation as
\begin{equation}
\begin{array}{l}
m_i n_0 [\frac{\partial{\mathbf{{v}_1}}}{\partial{t}}+\mathbf{v}_1.\nabla\mathbf{v}_0]  = e n_0 [\mathbf{E}_1 +(\mathbf{v}_1\times\mathbf{ B}_0)]  -\gamma_i K T_i  \nabla n_1+\gamma_i K T_i \frac{n_1}{n_0} \nabla n_0
\end{array}
\end{equation}
Here the resultant equation (Eq.(5)) contains two density gradient terms $\nabla n_1$ and $\nabla n_0$. The $n_1$ is the perturbation in the density caused due to the propagation of \emph{em} waves; while assuming the perturbation as $exp[i(ky-\omega t)]$, $\nabla n_1$ is replaced with $ik n_1$. Regarding the term $\nabla n_0$; $n_0$ is the number density of ions, the plasma we consider is fully ionized and the gravitational drift is effective to all the ions, therefore we replace $\nabla n_0$ with $n_0^{'}\hat{x}$, is the number density of ions drifted per unit length. Now the Eq.(5) becomes, 
\begin{equation}
\begin{array}{l}
m_i\alpha \mathbf{v}_1 = ie [\mathbf{E}_1+(\mathbf{v}_1\times\mathbf{B}_0)] + k \gamma_i K T_i \frac{n_1}{n_0}\hat{y} + i \gamma_i K T_i  \frac{n_1}{n_0} \frac{n_0^{'}}{n_0} \hat{x}
\end{array}
\end{equation}
where $ \alpha = (\omega-k v_0)$ and $\nabla {{v}_0} = 0$. Equating the x and y components of Eq.(6), we obtain the components of perturbed velocities for ions as  
\begin{equation}
\begin{array}{l}
v_{ix}=-\frac{\Omega_i^2}{\varpi_i}\frac{E_1}{B_0} + i \frac{\Omega_i^2 v_{ti}^2}{\alpha \varpi_i}\frac{n_0{'}}{n_0}\frac{n_1}{n_0} + i k \frac{\Omega_i v_{ti}^2}{\varpi_i}\frac{n_0{'}}{n_0} + i \frac{v_{ti}^2}{\alpha}\frac{n_0{'}}{n_0}\frac{n_1}{n_0}
\end{array}
\end{equation}
\begin{equation}
\begin{array}{l}
v_{iy}=i \frac{{\Omega_i}\alpha}{\varpi_i}\frac{E_1}{B_0} + \frac{\Omega_i v_{ti}^2}{\varpi_i} \frac{n_0{'}}{n_0}\frac{n_1}{n_0} + \frac{\alpha v_{ti}^2}{\varpi_i} \frac{n_1}{n_0} 
\end{array}
\end{equation}
Similarly we can deduce the components of perturbed velocities for electrons as
\begin{equation}
\begin{array}{l}
v_{ex}=-\frac{\Omega_e^2}{\varpi_e}\frac{E_1}{B_0} + i \frac{\Omega_e^2 v_{te}^2}{\omega \varpi_e}\frac{n_0{'}}{n_0}\frac{n_1}{n_0} - i k \frac{\Omega_e v_{te}^2}{\varpi_e} \frac{n_0{'}}{n_0} + i \frac{v_{te}^2}{\omega}\frac{n_0{'}}{n_0}\frac{n_1}{n_0}
\end{array}
\end{equation}
\begin{equation}
\begin{array}{l}
v_{ey}= -i \frac{{\Omega_e}\omega}{\varpi_e}\frac{E_1}{B_0} - \frac{\Omega_e v_{te}^2}{\varpi_e} \frac{n_0{'}}{n_0}\frac{n_1}{n_0} + \frac{\alpha v_{te}^2}{\varpi_e} \frac{n_1}{n_0} 
\end{array}
\end{equation}
Where $\varpi_i = (\omega-k v_0)^2 - {\Omega_i}^2$, $\varpi_e = \omega^2 - \Omega_e^2 $, $ v_{te,i} = ({\frac{\gamma_{e,i}KT_{e,i}}{m_{e,i}}})^{\frac{1}{2}} $ is the thermal velocities of electron and ion respectively.
\begin{figure}
\centering
\includegraphics[height=6cm]{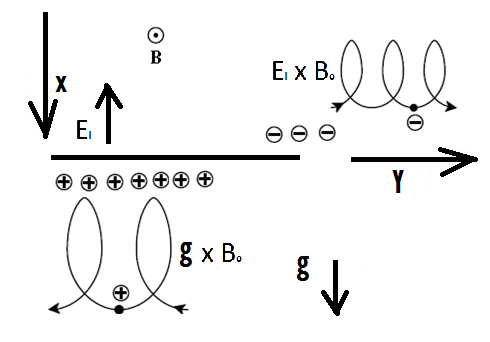}
\caption {Model of Gravity Induced Ion Acoustic Oscillation}
\end{figure}
Introducing the velocity components in perturbed continuity equation, we can derive a general dispersion relation valid for all range of frequencies. But the corresponding dispersion relation is very cumulative with numerous terms and analyzing it for the entire range of \emph{em} waves is very plausible. Therefore the dispersion relation is separated into two; \emph{em} waves of frequencies $\omega > \Omega_e$ and low frequencies $\omega <\Omega_e$.
For the high frequencies, the \emph{gief} is not effective, since all the terms containing the $E_1$ are vanishing out. This is an expected result, because $E_1$ is caused due to the drift of ions and therefore the possible interaction should be an ion cyclotron related one. For such a case, the maximum possible frequency will be the frequency of a coupled oscillation between the ions and electrons, i.e., \emph{lh} frequency which is less than $\Omega_e$. \\
For deducing the dispersion relation in low frequency range; by setting the condition $\omega < \Omega_e$ in the Eq.(7),(8),(9) \& (10), we get the velocity components in the following form,
\begin{equation}
\begin{array}{l}
v_{ix}=\frac{E_1}{B_0} - i k \frac{v_{ti}^2}{\Omega_i}\frac{n_1}{n_0}
\end{array}
\end{equation}
\begin{equation}
\begin{array}{l}
v_{iy}= -i \frac{(\omega-kv_0)}{\Omega_i}\frac{E_1}{B_0} - \frac{v_{ti}^2}{\Omega_i} \frac{n_0{'}}{n_0}\frac{n_1}{n_0}- k \frac{(\omega-k v_0)v_{ti}^2}{\Omega_i^2}\frac{n_1}{n_0}
\end{array}
\end{equation}
\begin{equation}
\begin{array}{l}
v_{ex}=\frac{E_1}{B_0} + i k \frac{v_{te}^2}{\Omega_e}\frac{n_1}{n_0}
\end{array}
\end{equation}
\begin{equation}
\begin{array}{l}
v_{ey}=  \frac{v_{te}^2}{\Omega_e} \frac{n_0{'}}{n_0}\frac{n_1}{n_0} - \frac{k \omega v_{te}^2}{\Omega_e^2} \frac{n_1}{n_0} 
\end{array}
\end{equation}
The perturbed equation of continuity for ions is given as 
\begin{equation}
\begin{array}{l}
(\omega-k v_0)\frac{n_1}{n_0} + i \frac{n_0^{'}}{n_0}v_{ix} - k  v_{iy}=0
\end{array}
\end{equation}
Substituting the values of $v_{ix}$ and $v_{iy}$ in the Eq.(15), we get
\begin{equation}
\begin{array}{l}
(\omega-k v_0)\frac{n_1}{n_0}+ i\frac{E_1}{B_0}[\frac{n_0^{'}}{n_0}+\frac{k(\omega-k v_0)}{\Omega_i}]+2k \frac{v_{ti}^2}{\Omega_i}\frac{n_1}{n_0}\frac{n_0^{'}}{n_0} + k^2 \frac{(\omega-k v_0)v_{ti}^2}{\Omega_i^2} \frac{n_1}{n_0} = 0
\end{array}
\end{equation}
Similarly, from the perturbed continuity equation of electron, we get
\begin{equation}
\begin{array}{l}
[\omega - \frac{2k v_{te}^2}{\Omega_e}\frac{{n_0}^{'}}{n_0} + \frac{ k^2\omega v_{te}^2}{\Omega_e^2}]  \frac{n_1}{n_0} + i\frac{E_1}{B_0}\frac{n_0^{'}}{{n_0}}= 0
\end{array}
\end{equation}
Substituting the value of $E_1$, from the Eq.(16) in to the Eq.(17), we can derive the dispersion relation as follows,
\begin{equation}
\begin{array}{l}
[1 + k^2 x_e^2 ] \omega^2 -[ k v_0 + k v + k v_{de} \frac{m_e}{m_i} + k^3 x_e^2 v_0] \omega \\+ [g \frac{n_0^{'}}{n_0}-2(\frac{n_0^{'}}{n_0})^{2} C_i^2 + k^2 v_0 v]=0
\end{array}
\end{equation}
where $ x_e = \frac{ v_{te}}{\Omega _{e}}$; $ v = (v_{di} - 2v_{de}) $; $ v_{di,e} = \pm \frac {\gamma_{i,e} KT_{i,e}}{eB} (\frac{{n^{'}_{\circ}}}{n_{\circ}})$, is the diamagnetic drift velocities; ${C_i} = [\frac{\gamma_i K T_i + \gamma_e K T_e}{m_{i}}]^{\frac{1}{2}}$, is the ion acoustic velocity.\\
\begin{figure}
\centering
\includegraphics[height=8cm]{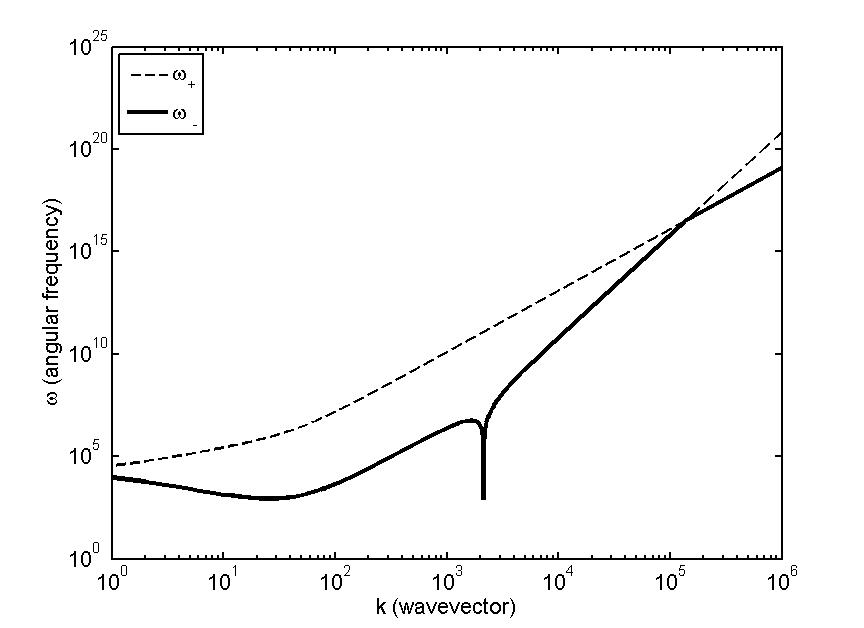}
\caption {Two modes of propagation of EM waves, for coronal plasma.($T_{e,i}=10^6$ $B=0.001T$, $g=274 ms^{-2}$, $n_0^{'}/n_0 =1$)}
\end{figure}
\section{RESULTS \& DISCUSSION} The above dispersion relation is a quadratic equation in $\omega $, but it is non linear in nature. There are two non linear terms in this relation, $k^3 {x_e}^2 v_{0}$ is in the coefficient of $\omega$ and $k^2 v_{0} v$ is in the coefficient of ${\omega}^0$. \textbf{ Generally a nonlinear term signifies the energy exchange process, therefore these two terms represent two physical processes related the energy exchange.} We find both these terms contain the wave vector ($k$) and gravitational drift velocity ($v_0$), which means that both these processes are happening due to the interaction between the gravity and the \emph{em} waves. \\
\subsection{Gravity induced ion acoustic waves}
There are several methods to analyze this dispersion relation, but we shall start in a simple way by analyzing the term of ion acoustic waves; this term is with the coefficient of the $\omega_0$ and also independent of wave vector. Therefore we can set the condition $k = 0$ (no perturbation) in the dispersion relation and it reduces to 
\begin{equation}
\begin{array}{l}
 \omega^2 + g \frac{n_0^{'}}{n_0}-2(\frac{n_0^{'}}{n_0})^{2} {C_i}^2 = 0
\end{array}
\end{equation}
For a high temperature plasma (solar corona), the term $g (n_0^{'}/n_0)$ is negligible, which leads to the solution $\omega = \pm \sqrt{2} {C_i} (\frac{n_0^{'}}{n_0})$. \emph{This shows that the electrons undergoes acoustic oscillations due to the inertial (gravitational) drift of ions, even when there is no perturbation.} This can be understood from fig.1 \& 2, in the absence of the perturbation, the direction of the electric field is in the -x axis and therefore the drift $\mathbf{E} \times \mathbf{B}$ will be in the y-axis. Since the gravitational drift $\mathbf{g} \times \mathbf{B} $ is in the -y axis, for ions these drifts are equal and opposite. But for the electrons the gravitational drift is negligible and hence they experiences the drift $\mathbf{E} \times \mathbf{B}$ in the y axis (see fig.2). The drift of electrons due to $\mathbf{E} \times \mathbf{B}$, causes a positive potential and by the inertia of the ions, these electrons overshoot back due to Coulomb's force, leads to the ion acoustic oscillations. As mentioned in the introduction, in the ultra vacuum condition the Coulomb's interaction is strong, therefore these ion acoustic oscillations are observable in the solar corona.\\
\begin{figure}
\centering
\includegraphics[height=8cm]{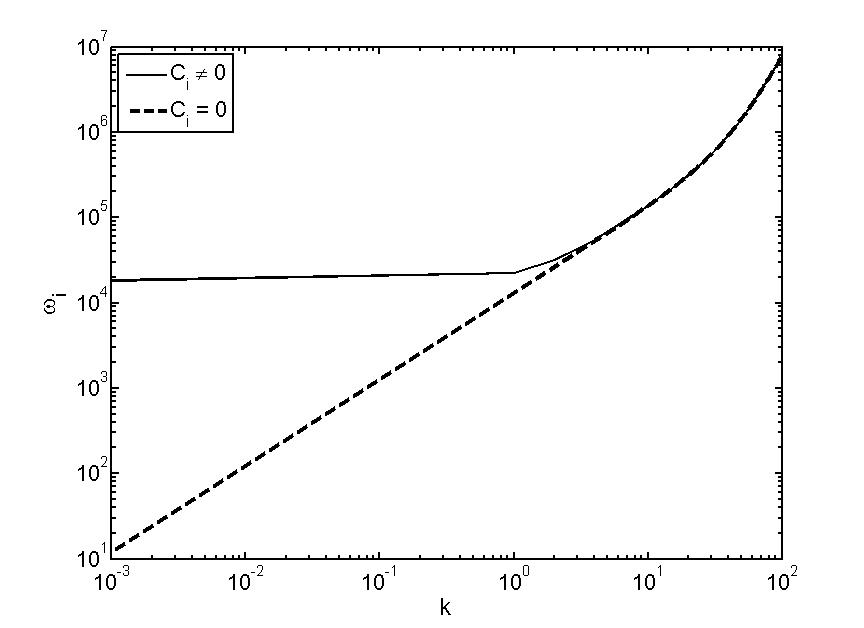}
\caption {Plot of growth rate for $ {C_i} \neq 0 $ \& $ {C_i} = 0$ ($ T_{e,i}=10^6 K$ $B=0.001T$, $g=274ms^{-2}$, $n_0^{'}/n_0 =1$)}
\end{figure}
The damping of ion acoustic waves is considered as one of the proposed mechanism for the solar coronal heating problem\cite{ref1,ref3,ref15}, therefore we should find out whether such damping occurs in our case. If we look in to the dispersion relation Eq.(18), we find the ion acoustic term is in the coefficient of ${\omega}^0$ with the nonlinear term $k^2 v_{0} v$.  This shows that certain energy sharing occurs between the \emph{em} waves and the ion acoustic waves. To confirm this, we have find out the unstable mode of \emph{em} waves from the dispersion relation, since the dispersion relation is a quadratic equation, the solution is given as ${\omega}_\pm = \frac{1}{2c}[-b\pm(b^2-4ac)]^{1/2}$, where a, b and c are the coefficients of $\omega^{2,1,0}$ respectively. To find the unstable mode of \emph{em} waves, the two roots ${\omega}_\pm$ are plotted (computationally using Mat-Lab programing) against the wave vector for the coronal plasma. Since one of the root is imaginary, the plot is done for the modulus of ${\omega}_{\pm}$ in a logarithmic scale to cover a wide range of frequencies, is shown in the fig.3. The real root (dashed line in the graph) represents the stable mode (undistorted part) of propagating \emph{em} waves and the imaginary root is the unstable mode (wave instability), where the effect of the nonlinear terms are reflected (solid line in the fig.3). 
\subsection{Instability \& Wave particle interactions} A wave instability means, the changes occurs in the physical properties of the wave due to the interaction with the local fields (wave-wave interaction) or the particles of the medium (wave-particle). During the wave-wave interaction; the propagating wave interacts with local oscillations such as Langmuir oscillations, acoustic oscillations and etc., becomes unstable. During this process, either of the wave grow at the cost of the other or if both the waves are damped, that will lead to the heating of the plasma.  In the wave particle interaction; the incoming wave interacts with the particles of the medium in such a way that, either the particles gain energy from the wave (Landau Damping process) or it looses to the wave (amplitude of the wave grow due to inverse Landau damping process). When the particles gain energy from the propagating waves, the waves are get damped (the group velocity of the wave, $v_g =\frac{d{\omega}}{dk}$, decreases) or the vice verse.\\
A simple way to study the instability is, finding the growth rate (imaginary part of $\omega$) of the instability and analyze it for the various parameters involved in the driving mechanisms. We have already found from the computational plot that one of the root is imaginary and the condition for a root to be imaginary is, $ 4ac >> b^2 $, therefore we can write the growth rate of the instability as,
$${\omega}_i =[\frac{c}{a} - \frac{b^2}{4a^2}]^{\frac{1}{2}}$$ 
\subsection{Damping of Ion acoustic waves}
Since we have started the discussion with ion acoustic waves, we shall see the effect ion acoustic waves over the unstable mode. The ion acoustic term is contained in the coefficient of $\omega_0$, we analyze the growth rate for very low wave vector range.  For smaller wave vector $k<1$, the coefficient $ a \approx 1 $ and  $\frac{b^2}{4} << c $ which gives  ${\omega}_i \approx \sqrt{c} $, i.e., the growth rate 
\begin{figure}
\centering
\includegraphics[height=8cm]{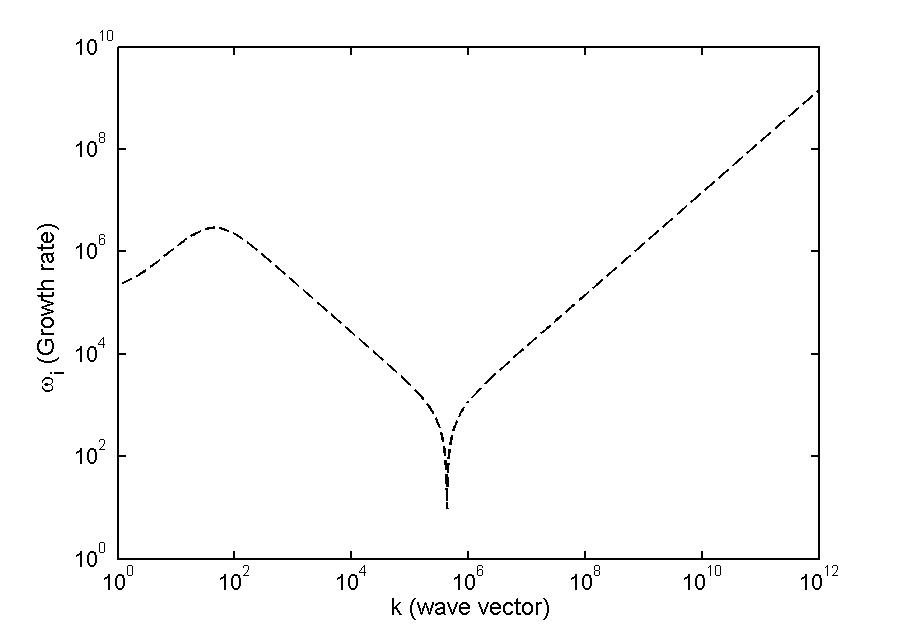}
\caption {Gravity Induced Resonant Emission (Plot of the growth rate for coronal plasma) ($T_{e,i}=10^6 K$ $B=0.001T$, $g=274 ms^{-2}$, $n_0^{'}/n_0 =1$)}
\end{figure} 
$$ \omega_i \approx [g \frac{n_0^{'}}{n_0}-2(\frac{n_0^{'}}{n_0})^{2} {C_i}^2 + k^2 v_{0} v]^{\frac{1}{2}}$$
For coronal plasma, the first term is negligible compared with ion acoustic term and hence the $\omega_i$ reduces to
$$\omega_i = [-2(\frac{n_0^{'}}{n_0})^{2} {C_i}^2 + \frac{k^2 v_{te}^2}{\omega_l^2} g (\frac{n_0^{'}}{n_0}) (2 + \frac{Ti}{Te})]^{\frac{1}{2}}$$
$ \omega_{lh}^2 = \left|\Omega_e \Omega_i \right|$. Now the growth rate contains two terms, the ion acoustic term  and the wave vector terms. Here the ion acoustic term is negative, which reduces the growth of the instability. When there is no perturbation($k\longrightarrow 0$),  $\omega$ becomes real which leads to the earlier results $\omega = \sqrt{2} {C_i}(\frac{n_0^{'}}{n_0})$, that the gravitational drift in the magnetized warm plasma generates ion acoustic waves.  Here we find in the absence of perturbation, the total thermal energy is converted in to the ion acoustic waves by the gravitational drift ions and the plasma becomes stable. But when ever there is a perturbation, the $v_{de}$couples with the $v_0$ of ions which results in the \emph{lh} oscillation, which leads to the damping of the ion acoustic waves. This could be more clear from the fig.2, in the perturbed plasma, the direction of $\mathbf{E_1}$ changes continuously at every point and the plasma becomes unstable. The growth of this unstable mode depends on the available energy; when $k$ increases, the $\omega_i$ increases at the cost of ion acoustic waves. This result is established in the fig.4, is a logarithmic plot between $\omega_i$ \& $k$, for $C_i \neq 0$ \& $C_i = 0$. 
In the fig.4, we find $\omega_i$ increases when $C_i = 0$ (dashed line, $v_g$ increases); this means that the unstable part of the \emph{em} wave grows at the cost of ion acoustic waves (wave-wave interaction). When $C_i \neq 0$, $\omega_i$ is constant ($v_g$ is constant), the \emph{em} wave does not grow. \textbf{Thus the propagation of \emph{em} waves in a magnetized warm gravitational plasma, causes the damping of ion acoustic waves.} The damping of ion acoustic waves have been observed in the solar corona \cite{ref21, ref2}, but this damping will not result in the heating of plasma; because, here the energy is transferred to another wave not to the particles. \\
\subsection{Gravity induced lower hybrid oscillations} 
Now, to find out the physics of the next nonlinear term, we are analyzing the unstable mode for the entire range of wave vector; the fig.5 shows the plot of $\omega_i$ with $k$, for a wide range of frequencies. 
\begin{figure}
\centering
\includegraphics[height=8cm]{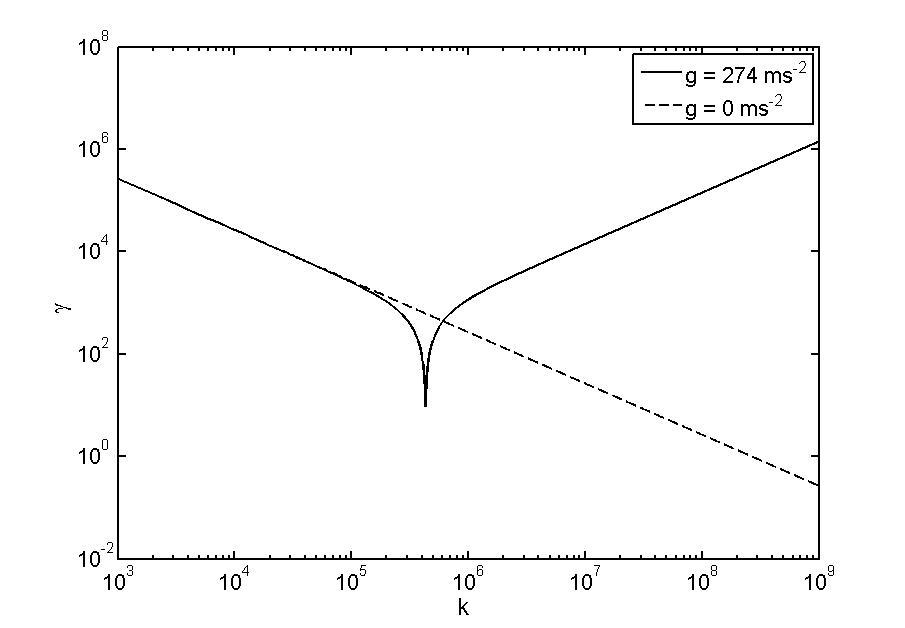}
\caption {Plot of growth rate for $g = 0 $ \& $g=274ms^{-2}$, ($ T_{e,i}= 10^6 K$, $B=0.001 T $, $n_0^{'}/n_0 =1$)}
\end{figure}
We find in this plot, the $\omega_i$ increases with $k$ up to the value of $10^2$, after that it decreases and then increases. The increase of $\omega_i$ in the low wave vector range; that we have already discussed, it is due to the damping of ion acoustic waves. But beyond the value $10^2$, the $\omega_i$ decreases with $k$, i.e., $v_g$ decreases, signifies that \emph{em} wave is losing the energy and at certain point, $v_g = 0$, where the propagating \emph{em} waves are completely absorbed by the particles. But, here after again $\omega_i$ increases, which reflects the resonant absorption and the re-emission of the \emph{em} waves by the ionized particles. This is a very interesting result from  the computational plot of the growth rate, now let shall analyze theoretically from the expression of $\omega_i$.
For high wave vector, $ k >> 1 $, the coefficient, $a \approx k^2 {x_e} ^2 $ , then
 $$\omega_i \approx [\frac{v_0 v_{de}}{2 {x_e}^2} - \frac{k^2 v_0^2}{4} - \frac{v_{de}^2}{4 k^2 {x_e}^4} ]^ {1/2}$$. 
Here we have neglected the cold plasma and the ion acoustic terms; at the resonant condition,  $v_g = 0$, by setting $ (\frac{\partial \omega_i}{\partial k})_{ \omega_{r}} = 0 $, we can derive the formula for the resonant frequency, $$ \omega_r^2 = \frac{ c^2 \omega_{l}^2}{g} \frac{n_0^{'}}{n_0} $$, where the $\omega_{l}^2 = \Omega_{e} \Omega_{i}$ is the lower hybrid frequency. \textbf{This shows that the perpendicular propagation of the \emph{em} waves, induces a \emph{lh} oscillation in the gravitational plasma, through which the ionized particles undergo the resonant absorption and re-emission.} This resonant absorption and re-emission is purely a gravity induced process (see fig.6), we shall name it as, the Gravity Induced Resonance Emission (\emph{gire}). The \emph{lh} resonance heating has been considered as a possible mechanism for the study of coronal emissions \cite{ref4,ref7,ref8,ref9,ref10,ref11,ref12,ref13,ref14,ref16,ref17,ref18, ref20}, therefore it is necessary to understand the physics of \emph{gire}.
\subsection{Gravity Induced Resonance Emission} To understand \emph{gire}, we should know the basic mechanism of \emph{lh} oscillation\cite{ref5}. The \emph{lh} oscillation is a coupled oscillation between the ions and electrons undergoing cyclotron motion, attached by Coulomb's force. In a magnetic field, all the ionized particles under go cyclotron motion, since the ions are inertial, the Lorentz force is less effective to the ions compared with electrons in a weak magnetic field. In such conditions, the ions are considered as less magnetic compared with electrons, which under go several rotation in the same period of ions. Therefore in a weak magnetic field, if there is a local electric field, the ions experiences more electrostatic force rather than the magnetic force. Thus the ions execute a back and forth motion, while the electrons are effective in the cyclotron motion (fig.7). Since the cyclotron motion of the ions and electrons are in opposite directions, in the every rotation they becomes closer and closer, get coupled by the Coulomb's Force and leads to a coupled oscillation under the magnetic field. Since this is a coupled oscillation between the ion and electron cyclotron motion and hence the frequency of this oscillation lies between these two ($\Omega_i > \omega_l > \Omega_e$). Now we know the basic requirement for the \emph{lh} oscillation is the presence of a local electric field, so that ions would be less effective to the magnetic field, undergo a back forth motion while the electrons are in the cyclotron motion. This is happening in our plasma, the \emph{gief} is the local electric field, so that the formation of \emph{lh} coupling is very effective in the coronal plasma. During this \emph{lh} oscillation, in each cycle of revolution, the particles are getting accelerated by the Coulomb's Force (fig.7a). The high mobility of electrons creates a deficit of field, for both the particles and hence they acquire the field energy from the incoming \emph{em} waves. This is a pure Landau damping process of \emph{em} waves induced by gravity, in which both the electrons and ions accelerated in the cyclotron motion to higher and higher energy level. The landau damping occurs in both the cycle of incoming wave, for the ions it happens at the upper half of the wave and for electrons in the lower half (see fig.7c). The electrons are lighter particles, undergo several rotations in the time period of ions and absorb more energy. This increases in the energy of electrons which results in the increase of it's Larmour radius, which gives more acceleration time for ion, result in the absorption of sufficient energy from the incoming radiation. This is just as a push and pull, the electrons are pushed by the \emph{em} waves which pulls the ions by Coulomb's force to the higher energy. When the ions are accelerated to higher energy level, they get ionized more to comply with the electrons to continue in the \emph{lh} oscillation (more work is needed to calculate total energy and acceleration time).     
Gravity is an inherent force and hence this process is continuous one, which is happening in the solar corona may be reason for the spectral emissions. Analyzing the spectral emissions on the basis of \emph{gire} is beyond the scope of this paper.\\
 \begin{figure}
\centering
\includegraphics[height=4.5cm]{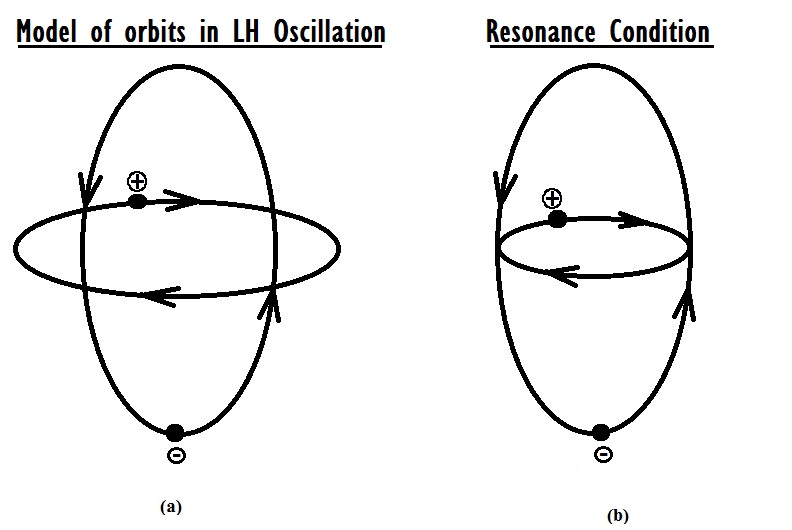}
\includegraphics[height=4.5cm]{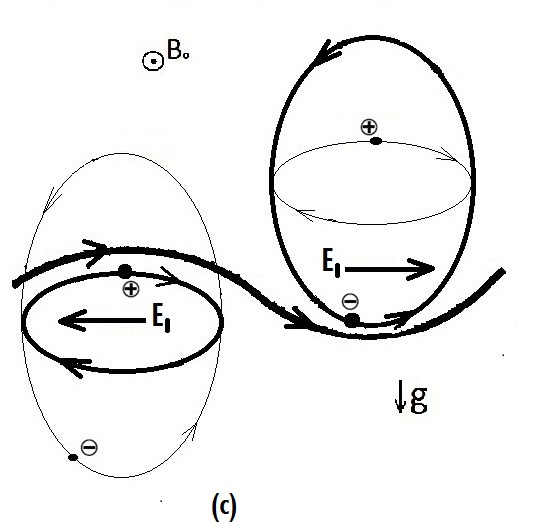}
\caption{Model orbits in \emph{lh} oscillation (a) for $\omega \ne \omega_{lh}$ (b) for $ \omega = \omega_{lh} $ (c) Photon absorption}
\end{figure}
\subsection{Theoretical Justification}
\begin{enumerate}
\item{For the study of motions perpendicular to B, the fluid theory is the best approximation}.
\item{We have used the MHD theory in its best representation for the study of warm plasma, i.e., by multiplying the number density $n_0$, with the single particle equation}.
\item{The Linearization method is a very simple and straight forward procedure for solving the differential equation}.
\item{The solar corona is best known ultra vacuum natural laboratory, is suitable for the study of particles interactions}.
\item{The plasma chosen for this study is a holistic one, including all the parameters}.
\item{And the formula we have derived is also a holistic one, convergent of all the fundamental quantities}.
\end{enumerate}
\section{Conclusion} In ultra vacuum condition, the fields are very effective and Coulomb's interactions are very strong and becomes a long range force. In such situation, the particle interactions are very dominant, the electric field induced by gravitational drift of ions is sufficient to cause the perturbations in the plasma. We have analyzed the effect of gravity over ionized particles, by deriving a general dispersion relation using MHD theory. We understand that the gravitational force couples with Lorentz force leads to the ion acoustic and lower hybrid oscillations. In the absence of \emph{em} perturbations, the gravity causes ion acoustic oscillations in the magnetized plasma and these waves are getting damped by the \emph{em} waves in the perpendicular direction of the magnetic field, which also results to the formation of lower hybrid oscillations. For higher wave vector, these gravity induced lower hybrid oscillation leads to the resonant absorption and re-emission of \emph{em} waves. We have analyzed these results for coronal plasma, but it is suitable to all low pressure plasmas. \\
\textbf{Acknowledgment}\\
The first author acknowledges the financial support by UGC under the UGC Research Award Scheme (2016-2018).

\end{document}